# Graphene Metal Adsorption as a Model Chemistry for Atmospheric Reactions

Y. Ortiz[1] and A.F. Jalbout[2*]

[1]*Instituto de Ciencias Físicas, UNAM, Cuernavaca, Morelos*

[2]*Departamento de Investigación en Física, Universidad de Sonora, Hermosillo, Sonora.*

*Abstract*

We propose a mechanism by which chloromethane and dichloromethane decomposition reaction occurs on the surfaces of graphene. To this end we have performed calculations on the graphene surface with metal adsorption on the sheet on the opposite side of reactions to reduce the formation of free-radical intermediates.

*Keywords:* Graphene: metal adsorption; DFT; Decomposition; Reaction

[*] To whom correspondence should be addressed
E-mail: drajalbout@gmail.com

**Introduction**

In recent work there has been considerable interest in the ability of nanotubes to alter certain chemical reactions using them as model chemistries for confined space. The nanotube confinement is believed to be important for reducing the energies needed for chemical reactions to take. Experimental work successfully has found that nanotubes may be used as reaction vessels for the $C_{60}O$ polymerization to yield a linear polymer $(C_{60}O)_n$ [1]. Another study showed that fullerenes may undergo chemical reactions inside of the nanotubes causing linear chains to form [2] as well as other catalytic reactions [3]. In other studies the paper by Halls et. al. revealed the impact of the presence of nanotubes on the Menshutkin SN2 reaction [4] as well as other related mechanisms [5-8].

The use of such research maybe in drug delivery [9] and a variety of other useful applications. What we have done in this study is to analyze the chemistry of decomposition mechanisms of chloromethanol, dichloromethanol and formyl chloride [10-13] on surfaces of graphene. It has been shown [14] that the adsorption of metals to the graphene surfaces can increase the potential reactivity of adsorbed molecules.

We believe that the decomposition of chlorinated hydrocarbons with and without the presence of water may be improved by the use of highly reactive radicals that may be obtained from hydrgon peroxide or ozone. If we can understand the way by which surfaces (i.e. graphene) can be used to improve the chemical nature of reactions this would be of importance.

As our strategy would be a mechanism to use graphene sheets to eliminate free radicals an understanding of such mechanisms might improve the negative impact of such products to living organisms. We will attempt to analyze different situations where metal adsorption to the surfaces of these graphene sheets improves the decomposition mechanisms of harmful reactions.

**Computational Methods**

In this work we have used the GAUSSIAN09 suite of packages [15]. We have used the B3LYP [16] method with the STO-3G optimization and 6-311++G** for single point energy calculations. The higher order calculations were performed to eliminate the BSSE error that may be intrinsic to systems of this type.

Per the previous references [12-13] we have considered the following reaction path:

$$CH_2\,(OH)Cl \rightarrow HCHO + HCl \quad (1)$$

$$HCHO + H_2O \rightarrow CH_2\,(OH)_2 \quad (2)$$

The latter equation represents the decomposition of dichloromethanol, that can also be described as:

$$CH\,(OH)\,Cl_2 \rightarrow ClCHO + HCl \quad (3)$$

$$ClCHO \rightarrow CO + HCl \quad (4)$$

The same reaction can proceed via a different pathway:

$$CH(OH)\,Cl_2 + H_2O \rightarrow ClCHO + H_2O + HCl \quad (5)$$

$$ClCHO + H_2O \rightarrow CH\,(Cl)\,(OH)_2 \quad (6)$$

$$CH(Cl)(OH)_2 + H_2O \rightarrow HCOOH + HCl + H_2O \quad (7)$$

As for the graphene surface we have considered a simple system of seven rings for the analysis. These represent pathways typical to the decomposition reactions and for purposes of discussion will be used as the basis for this manuscript. We have taken this pathway as an example for the efficacy of our proposed model. In future work we shall consider the effect of multiple Li adsorption to sheets of varying size.

**Results and Discussion**

In table 1 we display the relative energies (calculated with the B3LYP/6-311++G**//B3LYP/STO-3G method) of reactions (1-7) $\Delta E_I$: isolated chemical reactions 1-7, $\Delta E_{II}$: reactions 1-7 on the graphene surface and $\Delta E_{III}$: reactions 1-7 on the Li adsorbed on the graphene complex. Finally, for the sake of comparison $\Delta E_{IV}$ represents the MP2/6-31+G** energies of reaction using a confined zigzag (8,0) structure as the base chemistry [13]. The latter calculations were performed previously using the confined cavity of the nanotube for the chemical decompositions.

The corresponding molecular structures were optimized at the B3LYP/STO-3G level of the reactants on the Li+graphene surface are displayed in Figure 1 and the products on Figure 2. The Li atom slightly distorts off the center of the graphene surface a trend that is commonly observed in fullerene structures [17].

Due to space limitations we have not discussed alternative structures or the geometrical optimizations of the adsorption of the reactants/products on the pristine molecular surface.

*Reaction Pathway 1: $CH_2(OH)Cl \rightarrow HCHO + HCl$*

From table 1 is interesting to note the varying energies of the chemical decompositions. As we can see for the $CH_2$ (OH)Cl $\rightarrow$ HCHO + HCl (reaction 1) we obtain an energy of reaction that is significantly higher on the Li+graphene surface then the free reaction and the non-metal reaction. If we compare to the previous results using nanotube confinement [13] we observe that the calculations suggest that in the presence of such conditions such decompositions are in fact more reactive. While the effect is interesting, for many of these chemical reactions a higher barrier to reactivity is preferred as it minimizes the production of harmful free-radical intermediates.

As for the chemical structures the position and orientation of $CH_2(OH)Cl$ with respect to flake is similar to the case without Li. Measuring the angles between the co-linear carbon atoms of the flake we obtain a bend of around 3°.

For the structure of the products we obtain slightly different results. Compared to case without lithium atom the change is large, whereby molecules are almost out of the flake with a very different orientation. The distance of the lithium atom to the flake is similar in case we have just lithium atom over sheet.

*Reaction Pathway 2: $HCHO + H_2O \rightarrow CH_2\,(OH)_2$*

If we observe the second reaction $HCHO + H_2O \rightarrow CH_2\,(OH)_2$ the energy barriers using the tube are in the order of 13 kcal/mol which increase the barrier considerable compared to the isolated reaction. We observe that the reaction on the graphene surface has an energy of reaction of -13.48 kcal/mol. When the metal is adsorbed to the sheet we obtain a reaction energy of -17.25 kcal/mol. While these results are better for the nanotube case in the first initiation of the mechanism our energies of reaction will potentially limit the chemical reaction from proceeding to these end products.

Structurally, the reactants and products have interesting chemical formations. For the reactions compared with this case without Li the structures differ, by which the molecules are almost out of the flake. For the products the position and orientation of molecule $CH_2(OH)_2$ with respect to flake are similar to the non-Li case.

*Reaction Pathway 3: CH (OH) $Cl_2$ → ClCHO + HCl*

In the third mechanism CH (OH) $Cl_2$ → ClCHO + HCl our results using metal adsorption are not as notable as those of the nanotube surface. However, as it is a chain reaction from equation 1 the barrier is relatively high to prevent such molecules from forming.

The position and orientation of the reactant $CH(OH)Cl_2$ with respect to flake is similar in both cases. The angles between the co-linear carbon atoms of the flake yield a bend of $3^o$. We can consider qualitatively that the molecule is in the center of the flake. For the products the position and orientation of molecules with respect to flake are similar in the case without Li.

*Reaction Pathway 4: ClCHO → CO + HCl*

For this reaction the energies with the sheet are rather consistant and higher than the isolated mechanism. However, if we compare the efficacy of inhibiting the reaction on the surface when compared to the nanotube we observe a large difference. The value of the isolated reaction is 4.59 kcal/mol compared to 8.79 kcal/mol and -8.81 kcal/mol for the Li-graphene and nanotube case, respectively.

The molecule for the reactant in this reaction without Li is qualitatively on the center of the flake and separated from the flake by 4.1Å. It is interesting to note that the molecule have been shifted from the center of the flake. In the case of the reactants compared the case without Li there is a change and the molecules are nearly off the flake. Thus in certain cases it is useful to observe that adsorbing metal atoms may in fact case the systems to either adsorb to the surface or be removed when necessary.

*Reaction Pathway 5: CH(OH) $Cl_2$ + $H_2O$ → ClCHO + $H_2O$ + HCl*

For the fifth reaction CH(OH) $Cl_2$ + $H_2O$ → ClCHO + $H_2O$ + HCl our results are considerable improved relative to the nanotube case as we can see leading to a potential of 14.51 kcal/mol. This translates to the difficulty of the reaction upon metal placement on the surface. When we do not use a metal the reaction energies are rather high and make this decomposition quite feasible leading to a negative ecological impact. Structurally, the systems do not undergo drastic changes and remains near the flake. The distance of lithium atom to the flake is 1.9 Å and remain in the center of the ring.

*Reaction Pathway 6: ClCHO + $H_2O$ → CH (Cl) $(OH)_2$*

If we continue the reaction cascade to the next mechanism: ClCHO + $H_2O$ → CH (Cl) $(OH)_2$ reaction energies inside the nanotube are larger however from reaction 5 the difference from -25.23 kcal/mol to zero would mean that such a mechanism would have a very low probability of proceeding to reaction 6.

From a structural point of view, the reactants the case without Li there is a significant change in their arrangement, where the molecules are almost out of the flake with a very different orientation. The distance of lithium atom to the flake is 1.90 Å and remains in the center of the ring.

With regards to the products the molecules for this case without lithium atom is qualitatively on the center of the flake and with the presence of the metal atom they are shifted from the center.

*Reaction Pathway 7:* $CH(Cl)(OH)_2 + H_2O \rightarrow HCOOH + HCl + H_2O$

For the last reaction our results on the graphene surface are still much less favorable than inside the nanotube. This translates to the fact that the mechanism on the sheet will be less likely to take place than inside the tube or in an isolated case. The general trend observed for this pathway in both the structures of the reactants and products is that the metal is needed for the molecules to remain centered on the flake.

**Conclusions**

In this work we have explored the energies of reaction for the use of graphene to perform chemical decomposition of a chemical reaction pathway. It is important to note that when using the metals different model chemistries are obtained. This manuscript should serve as a basic framework for future experimentation in the field of atmospheric decomposition inhibitors using metal adsorption on graphene surfaces to reduce the rate at which specific chemical reactions take place.

It is important to make note of the fact that many of these chemical reactions have improved energy barriers for the decomposition mechanisms with respect to the nanotubes. This may translate to the improved ability of certain graphene structures to act as environmental buffers in harmful chemical reactions. With the use of metal adsorption to the surfaces we are able to minimize reaction energies in situations where we want to suppress harmful chemical reactions.

This work helps to support the fact that the graphene/organic layer may be readily susceptible to chemical reaction if it is induced to an external potential [18]. We have induced such a potential using Li adsorption to the surface of a simple graphene. Through this adsorption we are able to manipulate and potentially control certain chemical mechanisms.

This remains an interesting experimental question to be resolved and whether such implementation in order to control chemical reactions in the atmosphere can be performed. We believe through our calculations that while in certain situations confinement inside of nanotubes leads to improved results our graphene metal adsorption model chemistry acts to suppress important initial reactions that would lead to reduced production of harmful intermediates.

Additionally, there are implemented strategies for graphene adsorption whereas it is not yet clear how to control reactions inside of nanotubes. On a side note, we should mention that a calculation was performed using double metal adsorption to the sheet for reactaion pathway 7. For this system the reaction energies actually become more exothermic (-7.51 kcal/mol at the B3LYP/6-311++G** level of theory). As a potential useful application we can use the concentration ratio of Li atoms to control the outcome of specific chemical mechanisms.

**Acknowledgement**

We would like to thank VixTrend, LLC for valuable technical support in the development of this work. The authors also extend gratitude to Dr. Thomas Seligman for important discussions.

## Table and Figure Captions

**Table 1.** Relative energies (calculated with the B3LYP/6-311++G**//B3LYP/STO-3G method) of the different chemical reactions in kcal/mol whereby $\Delta E_I$: isolated chemical reactions 1-7, $\Delta E_{II}$: reactions 1-7 on the graphene surface and $\Delta E_{III}$: reactions 1-7 on the Li-graphene complex and $\Delta E_{Iv}$: reactions 1-7 using the MP2/6-31+G* zigzag (8,0) nanotube [13] for a model of the decompositions in a confined space.

**Figure 1**. Graphical depictions of optimized (B3LYP/STO-3G level) reactants (where the numbers correspond to the reactions described in the text) on the surface of the Li+graphene complexes, whereby bond lengths are in angstroms (Å) and angles are in degrees (°).

**Figure 2.** Graphical depictions of optimized products (B3LYP/STO-3G level) on the surface of the Li+graphene complexes (the numbers correspond to the reactions described in the text), whereby bond lengths are in angstroms (Å) and angles are in degrees (°).

| Chemical Reactions | No. | $\Delta E_I$ | $\Delta E_{II}$ | $\Delta E_{III}$ | $\Delta E_{iv}$ |
|---|---|---|---|---|---|
| $CH_2$ (OH)Cl → HCHO + HCl | 1 | 6.75 | 13.56 | 14.88 | -15.51 |
| HCHO + $H_2O$ → $CH_2$ (OH)$_2$ | 2 | -8.76 | -13.48 | -17.25 | 12.95 |
| CH (OH) $Cl_2$ → ClCHO + HCl | 3 | -5.06 | -6.33 | -6.76 | 0.00 |
| ClCHO → CO + HCl | 4 | 4.59 | 8.46 | 8.79 | -8.81 |
| CH(OH) $Cl_2$ + $H_2O$ → ClCHO + $H_2O$ + HCl | 5 | -5.05 | -20.41 | 14.51 | -25.23 |
| ClCHO + $H_2O$ → CH (Cl) (OH)$_2$ | 6 | -5.05 | 4.10 | 1.13 | 44.43 |
| CH(Cl)(OH)$_2$ + $H_2O$ → HCOOH + HCl + $H_2O$ | 7 | -1.76 | 1.02 | -4.74 | -24.77 |

**Table 1.**

**Figure 1.**

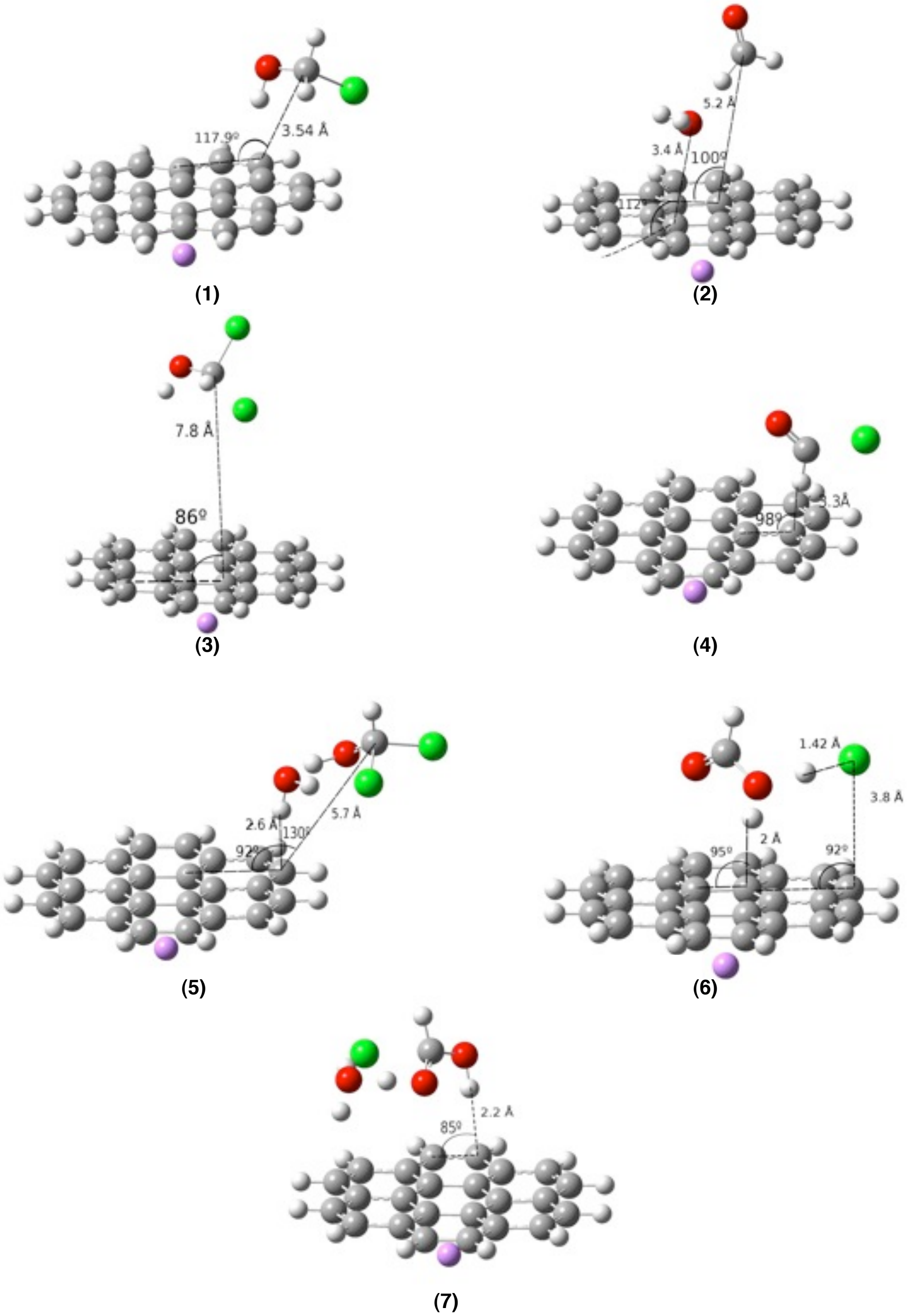

**Figure 2.**

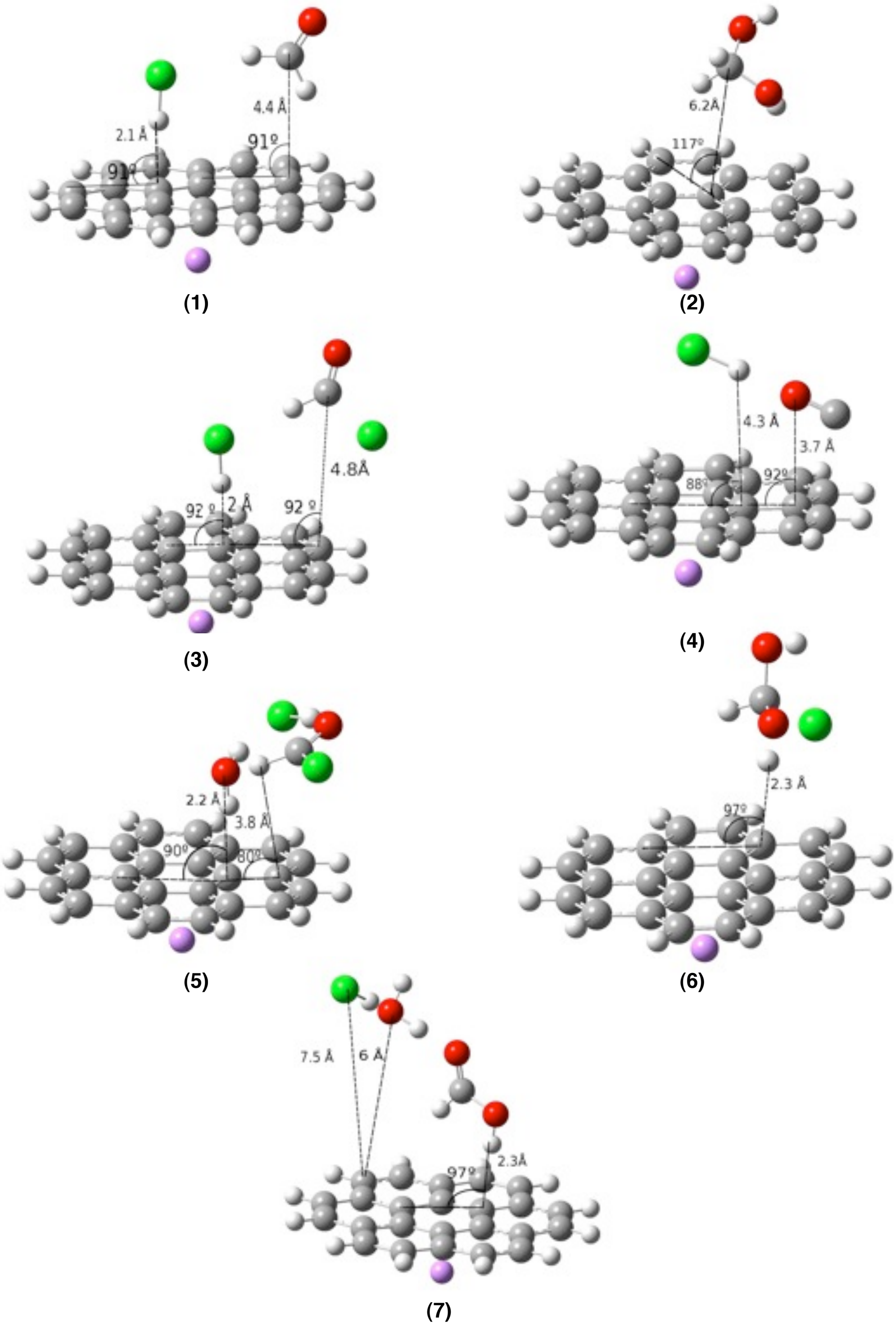